\documentclass[aps,twocolumn,amsmath,amssymb,superscriptaddress,prb]{revtex4}

\usepackage{graphicx}
\usepackage{color}
\usepackage{dcolumn} 
\usepackage{bm}      
\usepackage{mathrsfs}
\usepackage{framed}
\usepackage{float}
\definecolor{boxcolor}{rgb}{0.92,0.92,0.92}
\definecolor{commentcolor}{rgb}{0.19,0.80,0.19}
\newcommand{\diff}{\mathrm{d}}

\usepackage{textcomp}
\usepackage{subfigure}
\graphicspath{{figures/}}

\begin{document}
\title{NMR Technique for Determining the Depth of Shallow Nitrogen-Vacancy Centers in Diamond}

\author{Linh M. Pham}
\affiliation{Harvard-Smithsonian Center for Astrophysics, 60 Garden St., Cambridge, MA 02138, USA.}

\author{Stephen J. DeVience}
\affiliation{Department of Chemistry and Chemical Biology, Harvard University, 12 Oxford St., Cambridge, MA 02138, USA.}


\author{Francesco Casola}
\affiliation{Harvard-Smithsonian Center for Astrophysics, 60 Garden St., Cambridge, MA 02138, USA.}

\author{Igor Lovchinsky}
\affiliation{Department of Physics, Harvard University, 17 Oxford St., Cambridge, MA 02138, USA.}

\author{Alexander O. Sushkov}
\affiliation{Department of Chemistry and Chemical Biology, Harvard University, 12 Oxford St., Cambridge, MA 02138, USA.}
\affiliation{Department of Physics, Harvard University, 17 Oxford St., Cambridge, MA 02138, USA.}

\author{Eric Bersin}
\affiliation{Department of Physics, Harvard University, 17 Oxford St., Cambridge, MA 02138, USA.}

\author{Junghyun Lee}
\affiliation{Department of Physics, Massachusetts Institute of Technology, 77 Massachusetts Ave, Cambridge, MA 02139, USA.}

\author{Elana Urbach}
\affiliation{Department of Physics, Harvard University, 17 Oxford St., Cambridge, MA 02138, USA.}

\author{Paola Cappellaro}
\affiliation{Department of Nuclear Science and Engineering, Massachusetts Institute of Technology, 77 Massachusetts Ave, Cambridge, MA 02139, USA.}

\author{Hongkun Park}
\affiliation{Department of Chemistry and Chemical Biology, Harvard University, 12 Oxford St., Cambridge, MA 02138, USA.}
\affiliation{Department of Physics, Harvard University, 17 Oxford St., Cambridge, MA 02138, USA.}
\affiliation{Center for Brain Science, Harvard University, 52 Oxford St., Cambridge, MA 02138, USA.}

\author{Amir Yacoby}
\affiliation{School of Engineering and Applied Sciences, Harvard University, 15 Oxford St., Cambridge, MA 02138, USA.}
\affiliation{Department of Physics, Harvard University, 17 Oxford St., Cambridge, MA 02138, USA.}

\author{Mikhail Lukin}
\affiliation{Department of Physics, Harvard University, 17 Oxford St., Cambridge, MA 02138, USA.}

\author{Ronald L. Walsworth}
\email{rwalsworth@cfa.harvard.edu}
\thanks{Corresponding author}
\affiliation{Harvard-Smithsonian Center for Astrophysics, 60 Garden St., Cambridge, MA 02138, USA.}
\affiliation{Department of Physics, Harvard University, 17 Oxford St., Cambridge, MA 02138, USA.}
\affiliation{Center for Brain Science, Harvard University, 52 Oxford St., Cambridge, MA 02138, USA.}

\begin{abstract}
We demonstrate a robust experimental method for determining the depth of individual shallow Nitrogen-Vacancy (NV) centers in diamond with $\sim1$ nm uncertainty. We use a confocal microscope to observe single NV centers and detect the proton nuclear magnetic resonance (NMR) signal produced by objective immersion oil, which has well understood nuclear spin properties, on the diamond surface. We determine the NV center depth by analyzing the NV NMR data using a model that describes the interaction of a single NV center with the statistically-polarized proton spin bath. We repeat this procedure for a large number of individual, shallow NV centers and compare the resulting NV depths to the mean value expected from simulations of the ion implantation process used to create the NV centers, with reasonable agreement.
\end{abstract}

\maketitle

\section{Introduction}

\begin{figure}[]
  \includegraphics[width=0.95\columnwidth]{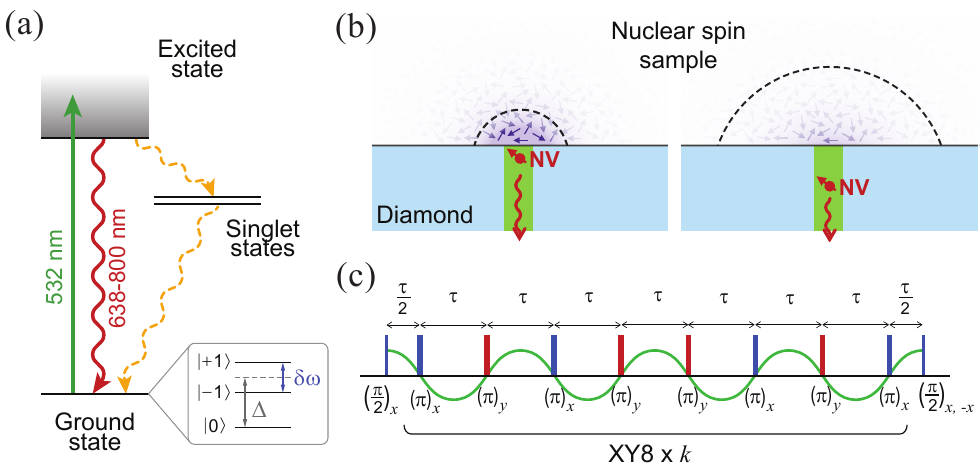}
  \caption{{\bf NV NMR Experiment.} (a) NV electronic energy level structure. (b) A confocal microscope addresses a single shallow NV center, which detects NMR signals from a few-nanometer region of sample on the diamond surface. Due to dipolar coupling, a shallow NV center (left) experiences a significantly stronger magnetic field from a smaller nuclear spin sample volume than a deep NV center experiences (right). The strength of the magnetic field at the NV center is indicated by the opacity of the nuclear spin sample, and the dashed lines qualitatively illustrate the volume of nuclear spin sample that contributes most of the NMR signal. (c) Larmor precessing statistically-polarized nuclear spins in the sample produce an effective AC magnetic field (green) that is detected by the NV sensor in a frequency-selective manner using an XY8$k$ pulse sequence.}
\label{fig:exp}
\end{figure}

The Nitrogen-Vacancy (NV) center in diamond is a leading platform for wide-ranging applications in sensing, imaging, and quantum information processing \cite{Taylor2008,MazeNature2008,BalasubramanianNat2008,DuttScience2007,NeumannNatPhys2010}. Key enabling properties of NV centers include exceptionally long electronic spin coherence times ($\rm{T_2}\gtrsim100\mu$s) \cite{Taylor2008, Doherty2013} and optical polarization and readout of the spin state (Fig.\ \ref{fig:exp}a) \cite{Doherty2013} in an atomic sized defect within the diamond crystal under ambient conditions.

Shallow NV centers within several nanometers of the diamond surface are especially useful for applications that rely on the strong dipolar coupling afforded by bringing the NV spin into close proximity to an external spin of interest. For example, quantum sensing \cite{Ajoy2015} and computing \cite{Cai2013} schemes in which NV centers are employed to control and read out the states of nuclear spins in samples tethered to the diamond surface require minimal separation between the NV and nuclear spins for strong coupling. In magnetic sensing applications, shallow NV centers with few nanometer separation from the magnetic field source have significant advantages over deeper NV centers and other magnetometers (e.g., SQUIDs) with much larger stand-off distances. Due to their close proximity to the sample, shallow NV centers (i) experience a larger magnetic field (i.e., dipolar fields fall off as $1/r^3$) and (ii) enable spatial resolution on a length-scale comparable to the stand-off distance, e.g., using scanning \cite{MaletinskyNNano2012,Grinolds2014}, super-resolution optical \cite{Maurer2010}, or Fourier imaging \cite{AraiArXiv2014} techniques. In particular, shallow NV centers have recently been used for nuclear magnetic resonance (NMR) spectroscopy and magnetic resonance imaging of nanoscale samples \cite{Mamin2013,Staudacher2013,DeVience2015} including single proton NMR and MRI \cite{Sushkov2014}.


Such applications of shallow NV centers depend crucially on accurate determination of the NV center depth, with uncertainty $\sim1$ nm. Shallow NV centers are most commonly formed via nitrogen ion implantation, with the NV center depth estimated using the Stopping and Range of Ions in Matter (SRIM) Monte-Carlo simulation\cite{ZieglerSRIM2008}. However, these estimates are statistical and thus do not provide the depth of any individual NV center. Furthermore, the simulations do not take into account crystallographic effects such as ion channeling, leading to underestimation of the NV depth by as much as a factor of two\cite{Toyli2010}. NV depth has also been estimated using secondary ion mass spectroscopy (SIMS) of nitrogen ions after implantation. Unfortunately, SIMS has a minimum detection threshold ($\sim 3 \times 10^{14}\ \rm{^{15}N/cm^{3}}$) and cannot be used to estimate individual NV center depths\cite{Toyli2010}.

Recently, the depth of individual NV centers has been experimentally determined using two techniques requiring highly-specialized and delicate apparatus.
The first technique takes advantage of F\"{o}rster Resonance Energy Transfer (FRET), determining NV depth by observing the coupling of single NV centers and a sheet of graphene brought in close proximity with the diamond surface. Measuring the NV fluorescence intensity as a function of separation between the graphene and diamond surface until the two are in contact and fitting the data with a theoretical model, NV depth can be determined with sub-nanometer uncertainty\cite{Tisler2013}.
In the second technique, a single shallow NV is employed to image, with $\sim1$ nm vertical resolution, dark electron spins assumed to be located at the surface of the diamond sample. The dark spin imaging resolution and consequently the uncertainty in NV depth determination is ultimately limited by the applied magnetic field gradient, the mechanical stability of the apparatus, and the $T_2^*$ of the dark spin\cite{Grinolds2014}.

In this paper, we present a robust method for extracting individual NV center depth with $\sim1$ nm uncertainty that can be easily performed with a scanning confocal microscope. We derive and analyze a model that describes the interaction of a single shallow NV center with a statistically-polarized nuclear spin bath, such as a proton-containing sample on the diamond surface, and discuss the conditions of validity of this model. Fitting the single-NV-measured proton NMR signal produced by microscope objective immersion oil, which has well understood nuclear spin properties, to the model expression, we determine depths for a large number of individual shallow NV centers and compare the measured depths with those expected from SRIM simulations. Finally, we discuss further application of this model to perform characterization of both NV centers as well as unknown nuclear spin samples on the diamond surface. Note that the experiments, model, and analysis presented here are a more detailed treatment of similar approaches to determining NV depth outlined in Refs. \citenum{Mamin2013,Staudacher2013,DeVience2015,Loretz2014}.

\section{Methods}


In our experiments we study negatively-charged NV centers formed via low-energy, low-dosage nitrogen ion implantation and subsequent annealing (see details in Sec. III and Table \ref{tb:nvdepth}), such that individual NV centers can be interrogated with a confocal microscope. To determine the depth of an individual NV center, we apply immersion oil to the diamond surface and measure the variance of the fluctuating NMR magnetic field at the NV center using a dynamical decoupling pulse sequence. The NMR magnetic field is created by a statistically-polarized subset of the proximal protons in the immersion oil, as shown in Fig.\ \ref{fig:exp}b. The protons undergo Larmor precession with a frequency determined by the applied static magnetic field (150-1600 G), but with a phase and amplitude that varies with every repetition of the pulse sequence. Although the net magnetization of the proton spin ensemble over the timescale of the entire experiment is negligible at the temperature and static fields applied in this work, the variance is nonzero and is proportional to the density of the proton bath.

We use an XY8$k$ pulse sequence, shown in Fig.\ \ref{fig:exp}c, to measure individual Fourier components of the NMR magnetic signal. We first optically pump the NV center electronic spin into the $m_{s}=0$ magnetic sublevel and create a coherent superposition of the $m_{s}=0$ and $m_{s}=1$ sublevels using a microwave (MW) $\pi/2$-pulse. The NV spin then undergoes periodic intervals of free evolution and 180$^{\circ}$ phase flips driven by resonant MW pulses, after which a final MW $\pi/2$-pulse converts the accumulated phase into an NV spin state population difference. The NV spin free evolution is governed by the time-dependent component of the total external magnetic field, which includes contributions from the proton NMR signal produced by the immersion oil on the diamond. The net accumulated NV spin phase is only appreciable when the evolution time $\tau$ is close to half the proton Larmor period.

The accumulated NV spin phase is measured by two consecutive near-identical experiments that project the final NV spin state first onto the $m_{s}=0$ state (resulting in a measurement of NV fluorescence $F_0$) and then onto the $m_{s}=1$ state (resulting in a measurement of NV fluorescence $F_1$), with appropriate choice of the final $\pi/2$-pulse phase. In order to remove common-mode noise from laser fluctuations, the two fluorescence signals are normalized to give the signal contrast $S = [(F_0 - F_1)/(F_0 + F_1)]$.

Measuring the signal contrast over a range of free evolution times $\tau$ results in slowly decreasing signal contrast for larger $\tau$, due to NV spin decoherence, and a narrower dip in contrast for specific values of $\tau$, caused by the nuclear spin Larmor precession. The background decoherence can be fit to an exponential function and normalized out, leaving the normalized contrast $C(\tau)$ with only the narrower NMR-induced dip (shown in detail in the appendix). The shape of this dip, described by Equation \ref{eq:Contr}, is determined by the magnetic field fluctuations produced by the dense ensemble of nuclear spins in the immersion oil on the diamond surface, as well as by the filter function corresponding to the XY8$k$ dynamical decoupling pulse sequence:
\begin{equation}
C(\tau) \approx \exp \left[-\frac{2}{\pi^2} \gamma_e^2  B_{\rm{RMS}}^2 K(N\tau) \right]. \label{eq:Contr}
\end{equation}
(An in-depth derivation is presented in the appendix.) Here $\gamma_e \approx 1.76 \times 10^{11}$ $\rm{rad/s/T}$ is the electron gyromagnetic ratio, $B_{\rm{RMS}}$ is the RMS magnetic field signal produced at the Larmor frequency by the nuclear spins, $K(N\tau)$ is a functional which depends on the pulse sequence and the nuclear spin coherence time, and $N$ is the number of $\pi$-pulses, which are separated by the NV spin free precession time $\tau$. 
As shown in the appendix, for the simplest case of a semi-infinite layer of a homogeneous nuclear-spin-containing sample on the most commonly used $\{100\}$-oriented diamond surface, $B_{\rm{RMS}}$ is related to the NV depth $d_{\rm{NV}}$ below the diamond surface by
\begin{equation}
B_{\rm{RMS}}^2 = \rho \left( \frac{\mu_0 \hbar \gamma_n}{4 \pi} \right)^2  \left(\frac{5\pi}{96 d_{\rm{NV}}^3}   \right),
\label{eq:BRMS1}
\end{equation}
where $\rho$ is the nuclear spin number density and $\gamma_n$ is the nuclear spin gyromagnetic ratio (for protons $\gamma_n \approx 2.68 \times 10^{8}$ $\rm{rad/s/T}$). More general cases of arbitrary nuclear spin quantum number and other diamond surface orientations can be calculated as described in the appendix. If the nuclear spin dephasing time ($T_{2n}^{*}$) is assumed to be infinite, then the functional $K(N\tau)$ is given by
\begin{equation}
K(N \tau) \approx (N \tau)^2 \text{sinc}^2 \left[\frac{N \tau}{2} \left(\omega_L - \frac{\pi}{\tau} \right)  \right],
\label{eq:infT2integr}
\end{equation}
where $\omega_L$ is the nuclear Larmor frequency. However, spectral broadening of the NMR signal due to diffusion or a finite dephasing time can also be included as shown in the appendix, in which case, the functional $K(N \tau)$ is given by
\begin{widetext}
\begin{multline}
K(N\tau) \approx \frac{2 T_{2n}^{*2}}{\left[1 +
   T_{2n}^{*2} \left(\omega_L - \frac{\pi}{\tau}\right)^2\right]^2} \left\{e^{-\frac{N \tau}{
     T_{2n}^{*}}} \left[\left[1 - T_{2n}^{*2} \left(\omega_L - \frac{\pi}{\tau}\right)^2\right] \cos \left[
        N \tau \left(\omega_L - \frac{\pi}{\tau}\right)\right]\right.\right.\\ -
      \left.\left. 2 T_{2n}^{*} \left(\omega_L - \frac{\pi}{\tau}\right) \sin \left[
        N \tau \left(\omega_L - \frac{\pi}{\tau}\right)\right]\right]+
     \frac{N \tau}{T_{2n}^{*}} \left[1 + T_{2n}^{*2} \left(\omega_L - \frac{\pi}{\tau}\right)^2\right] +
      T_{2n}^{*2} \left(\omega_L - \frac{\pi}{\tau}\right)^2 - 1 \right\}.
\label{eq:finT2integr}
\end{multline}
\end{widetext}
For a sample with well-known nuclear spin number density $\rho$ (e.g., $\rho = 68 \pm 5\ \rm{nm^{-3}}$ for the Nikon Type NF immersion oil employed in this work, measured using a Varian Unity Inova500C NMR system), the only free parameters in the fit expression are the NV depth $d_{\rm{NV}}$, the Larmor frequency $\omega_L$, and the nuclear spin dephasing time $T_{2n}^{*}$. The confidence with which each of these parameters can be extracted from a fit of Equation \ref{eq:Contr} to NV NMR data is strongly dependent on both the probed NV center properties and the applied pulse sequence.

In the limit of infinite $T_{2n}^*$, the strength of the NMR signal dip is entirely determined by the NV depth and the measurement pulse sequence duration $T = N \tau$, varying inversely with the former and directly with the latter. That is, for a fixed pulse sequence duration, shallower NV centers produce stronger NMR signal dips while deeper NV centers produce weaker NMR signal dips. As a result, pulse sequences with longer durations are necessary to acquire a strong enough NMR signal dip to confidently extract a depth estimate from a deeper NV center. On a related note, the infinite $T_{2n}^*$ limit is only valid when the pulse sequence duration is significantly shorter than $T_{2n}^*$; for sufficiently long pulse sequence duration, the NV detection bandwidth becomes narrow enough that the broadening of the NMR signal dip due to nuclear diffusion and spin dephasing can be observed and $T_{2n}^*$ can be extracted using the form of the functional $K(N\tau)$ given by Eq.\ \eqref{eq:finT2integr}.
The pulse sequence duration is eventually limited by the coherence time $T_2$ of the NV spin, however, which places upper bounds on the depth of NV centers and $T_{2n}^*$ of nuclear spin samples that can be extracted with this analysis. Recent work indicates a strong dependence of the NV $T_2$ coherence time on the NV depth for shallow NV centers.\cite{Myers2014} Assuming a typical value of $T_2 \sim 1$ ms found in deep NV centers and standard optical collection efficiencies ($<10\%$) we estimate that NV depths up to 300 nm below the diamond surface can be measured using the present method.


\begin{figure*}[]
\centering
  \includegraphics[ width = 6.75in]{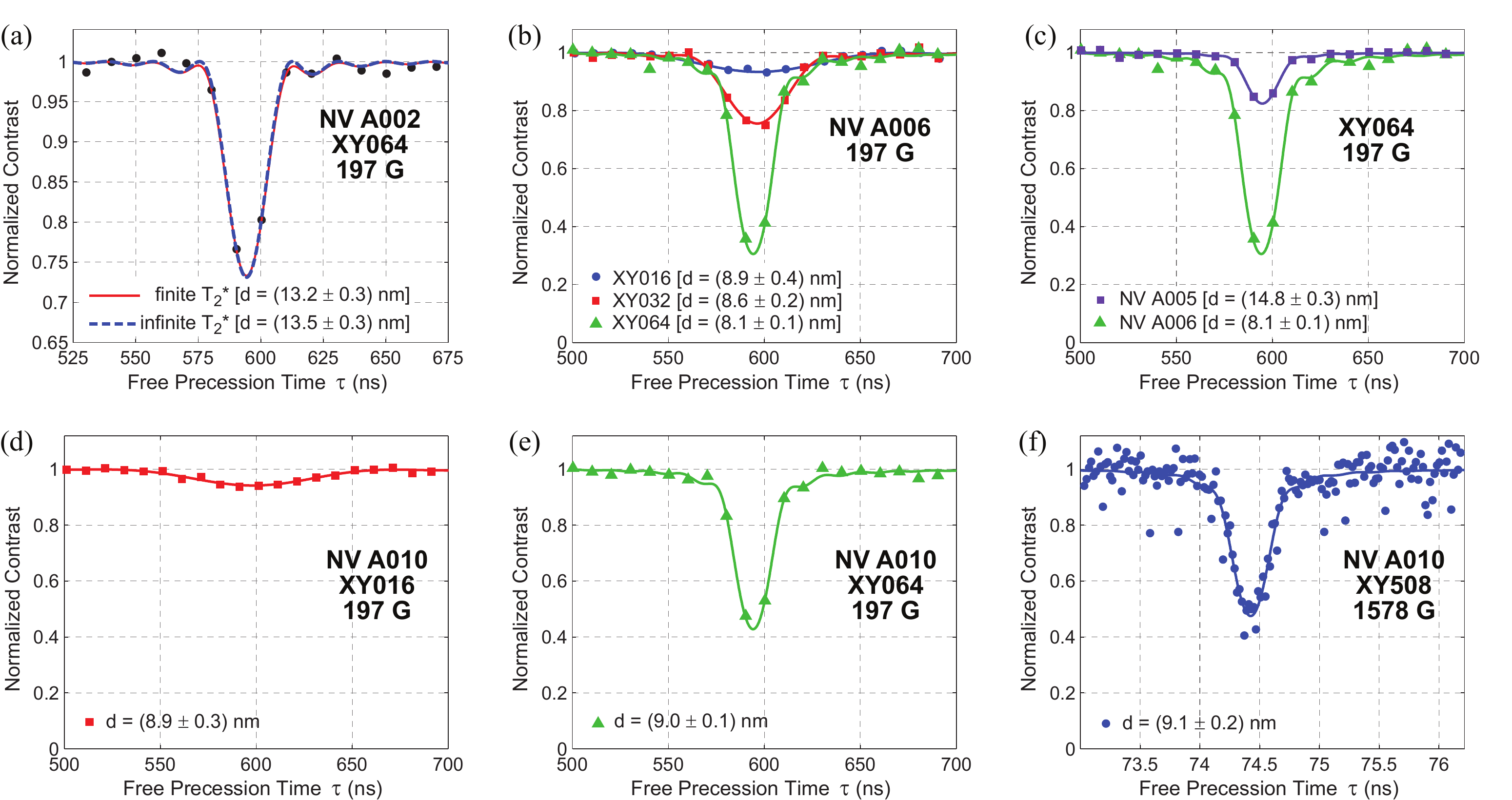}
  \caption{{\bf Example NV NMR proton spectra.} For all spectra, diamond sample and NV \#, pulse sequence, and applied static magnetic field are given in the bold inset label or in the symbol key, and the extracted NV depths are given in the symbol key. (a) NV NMR proton spectra data (black dots) measured with an XY064 pulse sequence at 197 G static field, analyzed assuming finite $T_{2n}^*$ (red solid curve) and infinite $T_{2n}^*$ (blue dashed curve). Both analyses fit the data well, with consistent NV depth values. (b) Proton NMR spectra measured with another NV center using different pulse sequences. The NV depths extracted from finite $T_{2n}^*$ fits (solid curves) are in reasonable agreement for all measurements. (c) Proton NMR spectra and finite $T_{2n}^*$ fits (solid curves) for two NV centers determined to have different depths under the same experimental conditions. The observed signal contrast dips vary strongly with NV depth. (d-f) Proton NMR spectra measured with the same NV center at different static field strengths and using different pulse sequences. Finite $T_{2n}^*$ fits (solid curves) yield consistent NV depths for all experimental conditions.}
\label{fig:infvfin}
\end{figure*}

\section{Results}
We performed measurements on 36 NV centers across 3 diamond samples, each synthesized via chemical vapor deposition (Element Six) and isotopically engineered to contain 99.999\% $^{12}$C. Sample A was implanted with 3-keV $^{15}$N$^{+}$ ions at a dose of $1 \times 10^9$ cm$^{-2}$; Sample B was implanted with 2-keV $^{15}$N$^{+}$ ions at a dose of $1 \times 10^9$ cm$^{-2}$; and Sample C was implanted with 2.5-keV $^{14}$N$^{+}$ ions with measurements taken in a region of 2D NV density $\sim 8 \times 10^7$ cm$^{-2}$. We employed a custom-built scanning confocal microscope to address single NV centers in each sample and fit the measured proton NMR signal from immersion oil on the diamond surface to Equation \ref{eq:Contr} in order to extract depth values for each NV center. 
A compilation of the measured properties of all the NV centers and diamond samples is given in Table \ref{tb:nvdepth}.
Proton spins in immersion oil have an expected $T_{2n}^* \sim 60\ \mu$s (corresponding to a linewidth $\sim 5$ kHz, see appendix for details) which is a longer nuclear $T_{2n}^*$ than can be extracted with the shallow NV centers used in the present work. Indeed, analysis of the measured NMR spectra data assuming infinite $T_{2n}^*$ (Eq.\ \ref{eq:infT2integr}) and finite $T_{2n}^*$ (Eq.\ \ref{eq:finT2integr}) generally give good agreement both in fits to the data and in NV depth extracted (Fig.\ \ref{fig:infvfin}a). However, since the infinite $T_{2n}^*$ condition does not hold strictly true for every measurement, we performed all analyses using the general case of finite nuclear $T_{2n}^*$, except where explicitly noted.

Figure \ref{fig:infvfin} shows typical measured proton NMR data from several representative NV centers in Sample A. The solid curves correspond to the best-fits of the model function to the data, from which NV depth estimates are extracted. We find that the contrast dip positions are in good agreement with those expected for the magnetic fields measured from the NV resonance frequencies, i.e., dips occur at $\tau = \pi/\omega_L$. Furthermore, we find that the fit expression yields consistent NV depth values even under different experimental conditions. For example, in Figure \ref{fig:infvfin}b, several measurements with different numbers of pulses were performed on the same NV center at the same static magnetic field. Fitting to each NMR spectrum independently, we extracted NV depth values that were in reasonable agreement with each other. Figures \ref{fig:infvfin}(d-f) show measurements and analyses of another NV center for which both the number of pulses and the static magnetic field were varied. Again, for all experimental conditions, the NV depth values extracted from the measurements are comparable to within their error bars. Figure \ref{fig:infvfin}(c) shows proton NMR data from two different NV centers measured with the same pulse sequence under the same experimental conditions (within the same diamond sample at the same static magnetic field) to illustrate the profound effect an NV center's depth can have on its sensitivity to NMR signals from nuclear spins at the diamond surface.

\begin{figure}[]
  \includegraphics[width=0.98\columnwidth]{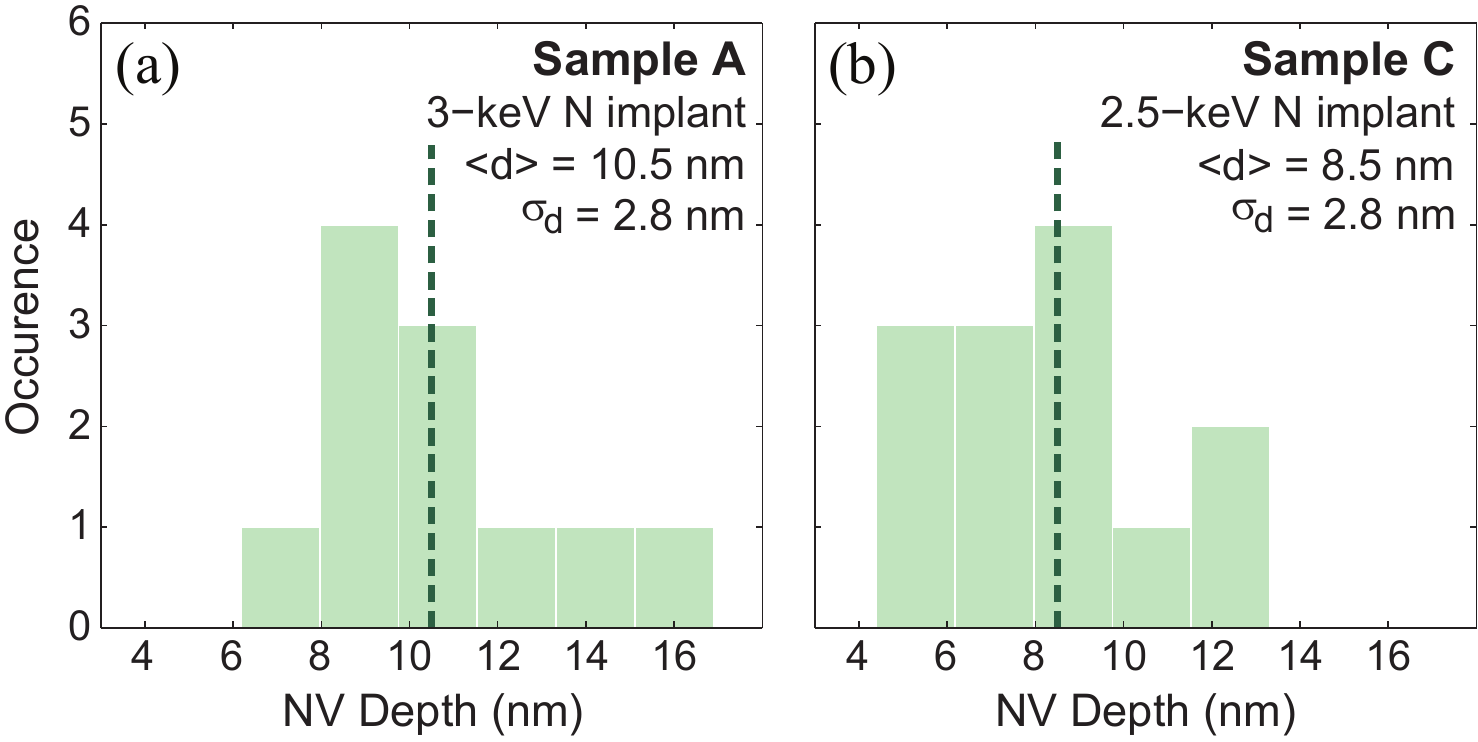}
  \caption{{\bf Histogram of measured NV depths in two diamond samples.} Estimated depths of (a) 11 NV centers in diamond Sample A, implanted with 3.0-keV $^{15}$N ions, and (b) 13 NV centers in diamond Sample C, implanted with 2.5-keV $^{14}$N ions. }
  \label{fig:histogram}
\end{figure}

Finally, we compared the distribution of NV depth values extracted from diamonds with different nitrogen implantation energies. Figure \ref{fig:histogram} shows histograms of the estimated depths for 11 NV centers in Sample A, which had been implanted with 3.0-keV $^{15}$N ions and 13 NV centers in Sample C, which had been implanted with 2.5-keV $^{14}$N ions (see also Table \ref{tb:nvdepth}). We found that the 3.0-keV implanted NV centers had a mean depth of 10.5 nm, with 2.8 nm standard deviation, and that the 2.5-keV implanted NV centers had a slightly shallower mean depth of 8.5 nm, with 2.8 nm standard deviation. In contrast, SRIM simulations predict a mean depth of ($5.2 \pm 2.1$) nm for 3.0-keV $^{14}N$ ion implantation and a mean depth of ($4.5 \pm 1.9$) nm for 2.5-keV; thus our measurements of NV depth are consistent with previous estimates that SRIM underestimates NV depth by as much as a factor of two.\cite{Toyli2010} However, it is important to note that the SRIM software estimates the distribution of implanted nitrogen ions whereas the NV NMR analysis estimates the depths of NV centers, which may have depth-dependent factors limiting their formation in diamond beyond the distribution of implanted nitrogen impurities. Furthermore, in addition to the NV centers whose extracted depths are represented in Figure \ref{fig:histogram}, in all diamond samples we observed that a fraction of the optically observed NV centers (e.g., roughly $1/2$ in Sample C) had optical and/or spin properties that were too unstable for any detailed measurements to be performed on them. These unstable optical and/or spin properties are likely symptomatic of very shallow NV centers whose depths cannot therefore be measured with the NMR technique presented in this paper. While this behavior may indicate a bias in the NV depth statistics extracted using this analysis technique, it also illustrates how this analysis may be applied towards determining how close to the diamond surface NV centers' optical and spin properties remain stable enough for sensitive spin measurements and furthermore provides an avenue for studying how surface treatments and processing can be used to stabilize very shallow NV centers. Both are topics of great importance in sensing, imaging, and quantum information applications that rely on shallow NV centers.

\begin{table}
\caption{\label{tb:nvdepth}Summary of the depths determined from 36 NV centers in 3 diamond samples under a range of external static field magnitudes $B_0$ and number of $\pi$-pulses $N$ used in the XY8$k$ measurement protocol. Sample A was implanted with 3.0-keV $^{15}$N ions; Sample B was implanted with 2.0-keV $^{15}$N ions; and Sample C was implanted with 2.5-keV $^{14}$N ions. In Samples A and C, measurements were performed on a random collection of NV centers such that the determined depth values reflect the NV depth distribution. In Sample B, measurements at 1609 G were performed only on NV centers that showed strong proton NMR signals for short averaging times; consequently these measurements are weighted towards shallower NV centers and do not accurately reflect the NV depth distribution.}
\begin{ruledtabular}
\begin{tabular}{ccccc}
Sample & NV \# & $B_0$ (G) & $\pi$-pulses & NV depth (nm)\\
\hline
A & 001 & 197       & 32              & 10.4(7) \\
A & 002 & 197       & 64              & 13.2(3) \\
A & 005 & 197       & 64              & 14.8(3) \\
A & 006 & 197       & 16, 32, 64      &  8.5(4) \\
A & 007 & 197       & 32, 64          &  9.0(4) \\
A & 008 & 197       & 64, 256         & 15.3(3) \\
A & 010 & 197, 1580 & 16, 32, 64, 508 &  8.9(5) \\
A & 012 & 197       & 32              &  8.3(3) \\
A & 104 & 150       & 16              &  6.4(2) \\
A & 110 & 150       & 64              & 10.7(4) \\
A & 111 & 150       & 64              & 10.0(2) \\
\hline
B & 009 & 206       & 64              & 10.7(7) \\
B & 022 & 159       & 32, 64, 96, 128 &  9.7(6) \\
B & 100 & 206       & 32              & 11(2)   \\
B & 112 & 1609      & 60              &  6.2(6) \\
B & 115 & 1609      & 124             &  7.7(3) \\
B & 116 & 1609      & 124             &  5.2(2) \\
B & 118 & 1609      & 124             &  6.5(3) \\
B & 119 & 1609      & 124             &  4.8(2) \\
B & 120 & 1609      & 124             &  4.8(2) \\
B & 121 & 1609      & 124             &  5.6(3) \\
B & 122 & 1609      & 124             &  5.0(2) \\
B & 123 & 1609      & 124             &  7.3(3) \\
\hline
C & 009 & 156       & 16, 32, 64, 96  &  8(1)   \\
C & 014 & 156       & 64              & 13.3(9) \\
C & 025 & 156       & 64              &  9.4(5) \\
C & 030 & 156       & 16              &  4.9(4) \\
C & 056 & 156       & 8, 16           &  4.7(2) \\
C & 075 & 156       & 64              &  7.4(2) \\
C & 090 & 156       & 64, 96, 128     &  7.5(5) \\
C & 093 & 156       & 64, 128         &  9.4(6) \\
C & 098 & 156       & 64, 96, 128     & 12(1)   \\
C & 107 & 156       & 64              &  8.6(4) \\
C & 111 & 156       & 16, 32          &  4.6(6) \\
C & 116 & 156       & 64              &  9.7(6) \\
C & 125 & 156       & 64, 128         & 11(1)   \\
\end{tabular}
\end{ruledtabular}
\end{table}

\section{Discussion}
Our robust NMR technique for determining the depth of shallow NV centers also enables detailed investigations of the effect of NV depth on other NV center properties. In particular, NV spin properties such as dephasing time $T_2^*$, coherence time $T_2$, and relaxation time $T_1$ may be characterized as a function of depth; furthermore, NV spectroscopic techniques may be applied to probe the local spin environment close to the diamond surface\cite{Bar-Gill2012}. Since magnetic sensing and quantum information applications that employ shallow NV centers also require long NV spin coherence times, better understanding and control of NV spin properties and the spin environment as a function of NV depth are key challenges.

In the present work, we applied our technique to determine NV center depth using a well-known nuclear sample. However, once an NV center's depth is determined, this information can be combined with our model to perform NV NMR studies of unknown nuclear samples. Also, as discussed in Section III, applying appropriate pulse sequences allows for the extraction of the nuclear spin $T_{2n}^*$, which can be used to study nuclear spin interactions and diffusion in the sample. Furthermore, by probing an unknown nuclear sample using multiple NV centers of differing depths, information about the nuclear spin distribution as a function of sample depth may be extracted.\cite{DeVience2015} 

\begin{acknowledgments}
This work was supported by DARPA (QuASAR program), MURI (QuISM program), NSF, and the Swiss National Science Foundation (SNSF). We gratefully acknowledge Fedor Jelezko for helpful technical discussions.
\end{acknowledgments}

\appendix
\renewcommand\thefigure{\thesection.\arabic{figure}}
\setcounter{figure}{0}
\section{NV Spin Decoherence Normalization}
\label{secA0:normalization}
\begin{figure}[]
  \includegraphics[width=0.98\columnwidth]{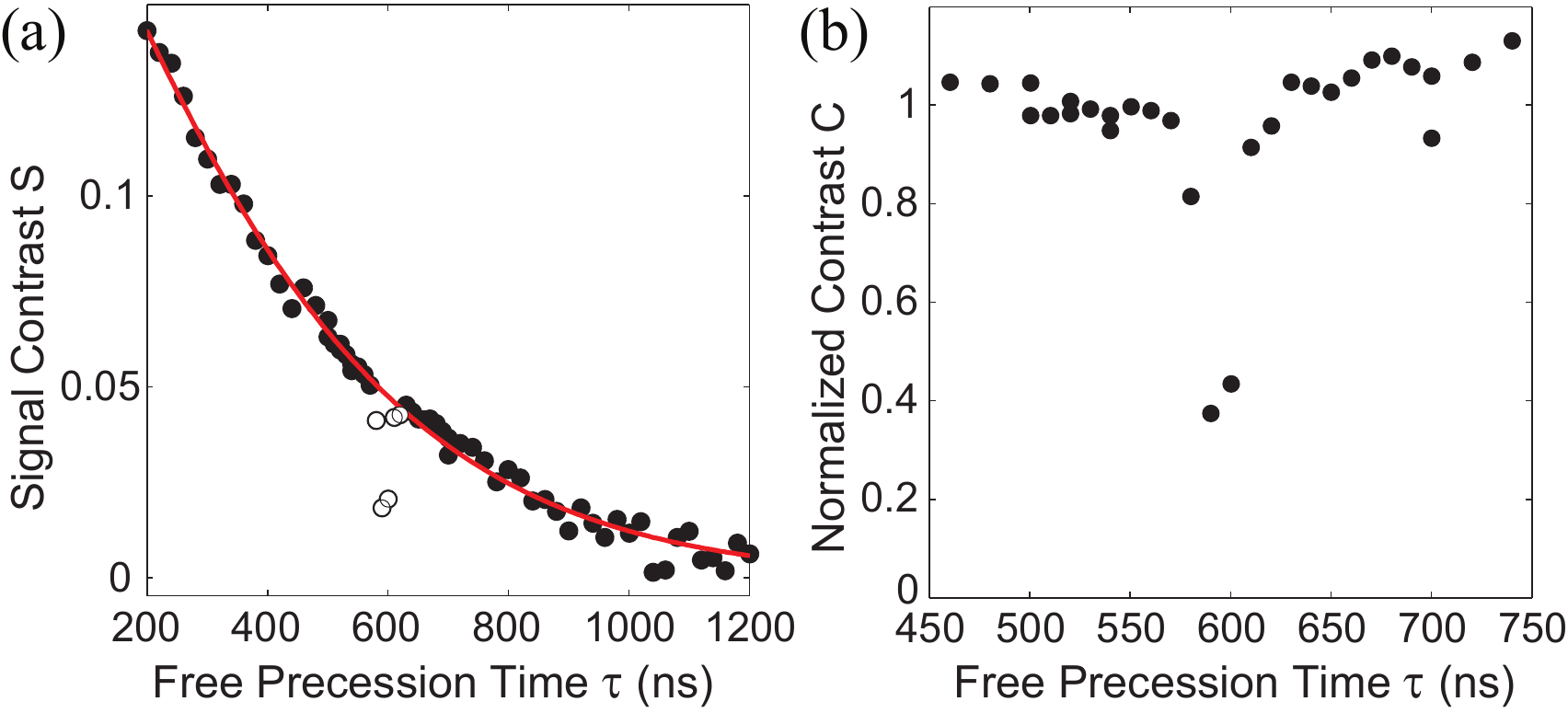}
  \caption{(a) Example NV signal contrast $S(\tau)$ data (circles) measured by applying an XY064 pulse sequence on NVA006 (Sample A). The decay due to NV spin decoherence is fit to a stretched exponential function (line), excluding the data which makes up the narrow NMR dip (open circles). (b) Normalized contrast $C(\tau)$ data isolates the NV NMR signal.}
\label{fig:decoherence}
\end{figure}

As described in the main text, two NV$^{-}$ spin-state-dependent fluorescence measurements $F_0(\tau)$ and $F_1(\tau)$ are acquired from consecutive, near-identical but independent dynamical decoupling experiments, each with $\pi$-pulses spaced by time $\tau$. For $F_0(\tau)$, the final $\pi/2$-pulse projects the NV spin coherence onto the $\vert 0 \rangle$ state, whereas for $F_1(\tau)$ the pulse phase is reversed to project the coherence onto $\vert \pm 1 \rangle$. This procedure removes common-mode noise from laser fluctuations occurring on timescales $\gtrsim\tau$. The fluorescence signals are described as a signal contrast, $S(\tau)$, of the form:
\begin{equation}
S(\tau) =\frac{F_0(\tau) - F_1(\tau)}{F_0(\tau) + F_1(\tau)}.
\end{equation}
The signal contrast effectively measures the projection of the NV spin coherence after the pulse sequence onto the coherence at the beginning of the sequence. Measuring $S$ over a range of free evolution times $\tau$ yields a slow decay due to NV spin decoherence and a narrow dip due to nuclear spin Larmor precession. The background NV spin decoherence can be fit to a stretched exponential function, excluding the data points which make up the narrow dip corresponding to the NMR signal, as shown in Fig. \ref{fig:decoherence}(a). Dividing by this exponential fit function yields a normalized contrast $C(\tau)$ which isolates the NMR signal in the NV measurement, as shown in Fig. \ref{fig:decoherence}(b).

\renewcommand\thefigure{\thesection.\arabic{figure}}
\setcounter{figure}{0}
\section{NV NMR Lineshape}
In this appendix, we present a derivation for the signal expected from an NV NMR measurement made with a dynamical decoupling sequence. We adopt the non-unitary Fourier transform in angular frequency units, such that the Fourier transform pair for $f(t)$ is defined as\cite{Bracewell}:
\begin{gather}
f(t) = \mathscr{F}^{-1} (f(\omega)) = \frac{1}{2 \pi} \int_{- \infty}^{+ \infty} f(\omega) e^{i \omega t} \diff \omega, \notag\\
f(\omega) = \mathscr{F}(f(t)) = \int_{- \infty}^{+ \infty} f(t) e^{-i \omega t} \diff t. \label{eq:conv}
\end{gather}
With the previous expression, Parseval's theorem reads as:
\begin{align}
& \int_{- \infty}^{+ \infty} f(t)g^*(t) \diff t = \frac{1}{2 \pi} \int_{- \infty}^{+ \infty} f(\omega) g^*(\omega) \diff \omega \notag\\ \rightarrow
& \int_{- \infty}^{+ \infty} |f(t)|^2 \diff t = \frac{1}{2 \pi} \int_{- \infty}^{+ \infty} |f(\omega)|^2 \diff \omega, \label{eq:eqFull}
\end{align}
and the expressions for the Dirac delta and convolution functions are:
\begin{gather}
\delta(\omega - \omega') = \frac{1}{2 \pi} \int_{- \infty}^{+ \infty} e^{i t (\omega - \omega')} \diff t \notag\\
\mathscr{F}(f \ast g) = f(\omega) g(\omega). \label{eq:convDel}
\end{gather}

\subsection{Signal from a Dynamical Decoupling Sequence}
\label{secA:comment}
During the dynamical decoupling measurement sequence, the NV spin coherence accumulates some phase $\Delta\phi(\tau)$ due to evolution in the presence of magnetic fields. In this work, the magnetic field of interest is the NMR signal from statistically-polarized spins in the sample on the diamond surface. After normalizing out contributions due to background NV spin decoherence (see Appendix A), the contrast is related to the accumulated phase by:
\begin{align}
&C(\tau) = \langle \cos(\Delta\phi(\tau)) \rangle. \label{eq:contr}
\end{align}
The brackets around $\cos(\Delta\phi(\tau))$ indicate that a typical fluorescence measurement is an average over many repeated, nominally identical dynamical decoupling experiments. If the accumulated phase $\Delta\phi(\tau)$ follows a normal distribution centered at zero with variance $\langle \Delta\phi^2(\tau) \rangle$ as will typically be the case for an NMR signal from a statistically-polarized nanoscale sample, then the average over the cosine can be converted to an exponential function of the variance using the relationship\cite{Recchia1996}:
\begin{align}
&\langle f(X) \rangle = \int_{-\infty}^{\infty} f(x)p(x) \diff x, \label{eq:Gaussint}
\end{align}
where $p(x)$ is the probability distribution function for random variable $X$. Applying the integral of Eq.\ \eqref{eq:Gaussint} to Eq.\ \eqref{eq:contr} yields:
\begin{align}
&C(\tau) = \exp(-\langle \Delta \phi^2(\tau) \rangle/2). \label{eq:contr1}
\end{align}

\begin{figure}[]
  \includegraphics[width=0.75\columnwidth]{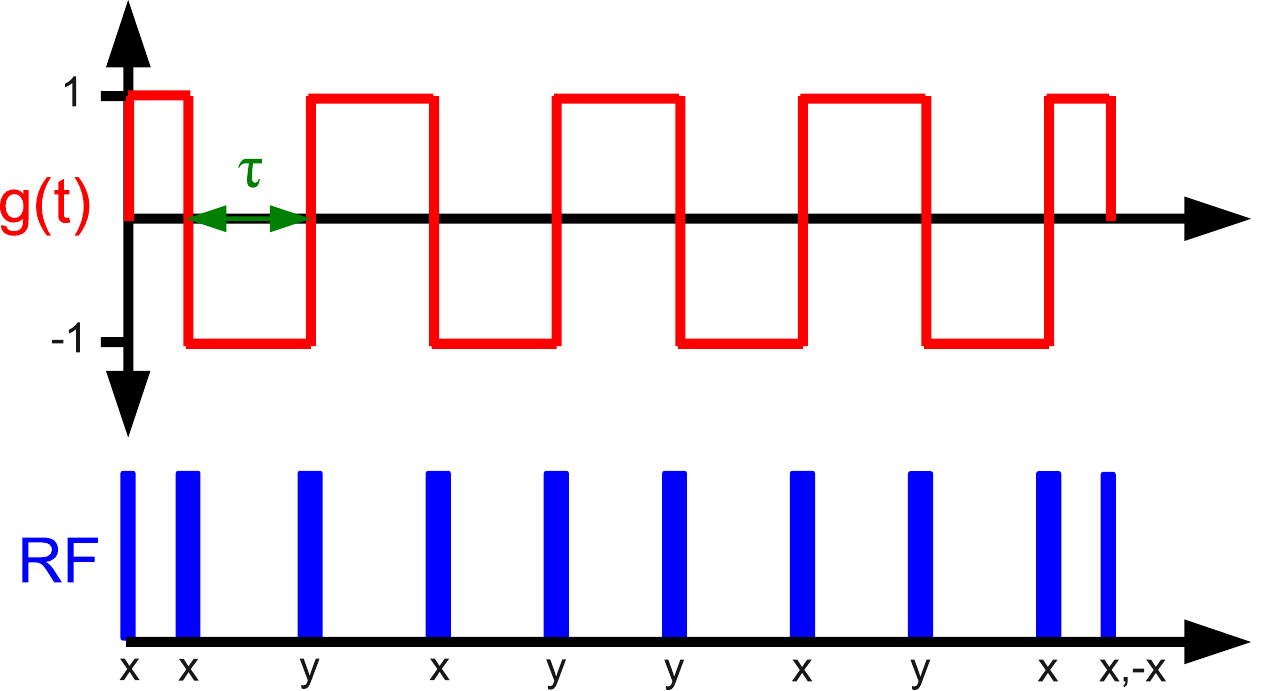}
  \caption{The dynamical decoupling sequence, induced by resonant MW pulses with phases as labeled, defines a function $g(t)$ describing the direction of NV spin precession in response to a magnetic signal $B_z(t)$.}
\label{fig:avg1}
\end{figure}

Phase is accumulated during the dynamical decoupling sequence as the NV electronic spins Larmor precess in the presence of a magnetic field signal $B_z(t)$, where $z$ is the NV quantization axis. (The NV spin Larmor precession from the static background field $B_0$ is removed by working in the rotating reference frame). The sign of phase accumulation (i.e., positive or negative phase accumulation) is reversed by each $\pi$-pulse of the sequence, and can be represented over time as a function $g(t)$, as shown in Fig.\ \ref{fig:avg1}. The total phase accumulated at the end of the sequence is then:
\begin{align}
&\Delta \phi(\tau) = \gamma_e \int_{- \infty}^{+ \infty} g(t) B_z(t) \diff t, \label{eq:phase}
\end{align}
where $\gamma_e$ is the gyromagnetic ratio for the NV electronic spin (in units of rad/s). The accumulated phase variance can be expressed in terms of a correlation function between measurements across times $t$ and $t'$:
\begin{align}
&\langle \Delta \phi^2(\tau) \rangle = \gamma_e^2 \langle \int_{- \infty}^{+ \infty} g(t) B_z(t) \diff t \int_{- \infty}^{+ \infty} g(t') B_z(t') \diff t' \rangle. \label{eq:corr1}
\end{align}
We now assume temporal translational invariance for the local and time-dependent field correlator:
\begin{equation}
\langle B_z(t) B_z(t') \rangle = S_B(t-t'). \label{eq:corr1Bz}
\end{equation}
Then we can write:
\begin{align}
&\langle \Delta \phi^2(\tau) \rangle = \gamma_{e}^2  \int_{-\infty}^{+\infty}\int_{-\infty}^{+\infty} S_B(t-t') g(t)g(t') \diff t \diff t'  \nonumber \\
& = \gamma_{e}^2  \int_{-\infty}^{+\infty}\int_{-\infty}^{+\infty} S_B(\tau) g(t')g(\tau + t') \diff \tau \diff t' \nonumber \\
& = \gamma_{e}^2  \int_{-\infty}^{+\infty}\int_{-\infty}^{+\infty} S_B(\tau) g(\tau + t') \diff \tau  g(t') \diff t' \nonumber \\
& = \gamma_{e}^2  \int_{-\infty}^{+\infty} J^z_{1,2}(t') g(t') \diff t' \nonumber \\
& = \frac{\gamma_{e}^2}{2 \pi}  \int_{-\infty}^{+\infty} J^z_{1,2}(\omega) g(\omega) \diff \omega,
\end{align}
where in the last line of the previous expression we have used Parseval's theorem. Since the term $J^z_{1,2}(t')$ is nothing but a convolution, one can easily conclude that:
\begin{align}
\langle \Delta \phi^2(\tau) \rangle &= \frac{\gamma_{e}^2}{2 \pi}  \int_{-\infty}^{+\infty} S_B(\omega) g(-\omega) g(\omega) \diff \omega \nonumber \\
&= \frac{\gamma_{e}^2}{2 \pi}  \int_{-\infty}^{+\infty} S_B(\omega) |g(\omega)|^2 \diff \omega.\label{eq:Stau13}
\end{align}
The quantity $S_B(\omega)$ represents the spectral density of the effective NV spin phase noise resulting from the magnetic field $B_z(t)$ and manipulation of the NV spin by repeated dynamical decoupling sequences; it can be computed as follows:
\begin{align}
S_B(\omega) &= \langle \left| B_{z}(\omega) \right|^2 \rangle \notag\\
&= \int_{- \infty}^{+ \infty} \langle B_{z}(0)B_{z}(t') \rangle e^{-i \omega t'} \diff t'. \label{eq1}
\end{align}

\subsection{Application to NMR Signals}
\label{secB:app1}

\subsubsection{Correlation Functions}
\label{secB:corr1}

We consider the NMR magnetic signal $B_{z}(t)$ originating from nuclear spins on the surface of the diamond and in the vicinity of a shallow NV center (see Fig. \ref{fig:geometry}). The statistically-polarized nuclear spin ensemble produces fluctuations in $B_z(t)$. For an ensemble of point dipoles, $B_{z}(t)$ at the NV center can be written as:

\begin{align}
B_{z}(t) =  \sum_{j} D_{j} \left[\right. &3 u^j_xu^j_z I_x^j(t) + 3 u^j_yu^j_z I_y^j(t) \notag\\
&\left.\qquad\ + (3u^j_zu^j_z-1) I_z^j(t) \right],
\label{eq:Bsum}
\end{align}

\noindent where the NV is coupled to many nuclear spins $j$ at positions given by a distance $r_j$ and a unit vector $u^j$ (which can be written in terms of of its coordinates $u^j_x, u^j_y, u^j_z$). The coupling factor is $D_j = (\mu_0 \hbar \gamma_n)/(4 \pi r_j^3)$, where $\gamma_n$ is the gyromagnetic ratio of the nuclei and $r_j$ is the distance between the NV center and nuclear spin $j$. Terms $I^j_{x,y,z}$ represent the operator projection of nuclear spin $j$ along the $x$, $y$, and $z$ axes.

Using Eq.\ \eqref{eq:Bsum}, the time-dependent correlator for the NMR magnetic field can be expressed as:
\begin{widetext}
\begin{align}
\langle B_{z}(0)B_{z}(t) \rangle= \langle &\sum_j  D_j(r_j) \left[  3u^j_xu^j_z I_x^j(0) + 3u^j_yu^j_z I^j_y(0) + (3u^j_zu^j_z-1) I^j_z(0) \right] \notag \\
& \sum_i D_i(r_i) \left[  3u^i_xu^i_z I^i_x(t) + 3u^i_yu^i_z I^i_y(t) + (3u^i_zu^i_z-1) I^i_z(t) \right] \rangle. \label{eq:fullCorr}
\end{align}
\end{widetext}

For an ensemble of nuclear spins that do not interact with each other, time-dependent correlators can be defined for every spin's operator projection along each of its axes:
\begin{equation}
\langle I_{\alpha}^j(0) I_{\beta}^i(t) \rangle = \delta_{\alpha,\beta} \delta^{i,j} f^{\alpha,\beta} (I,T,B_0,t). \label{eq:ParCorr}
\end{equation}
The function $f^{\alpha,\beta}$ represents the local nuclear spin-spin correlation function. By treating the nuclear spins as paramagnetic, the correlations between different nuclear sites are identically zero. Note that the correlator is a function of the nuclear spin's total spin quantum number $I$ as well as the temperature $T$ and the applied field $B_0$ (which determines the Larmor frequency of the nuclei). In the simple case in which the external magnetic field for the nuclei is applied along the NV axis one can write $f^{x,x} = f^{y,y}$, i.e., behavior in the transverse plane is independent of the relative phase between the nuclear spin and the NV. Moreover, all nuclear spins of the same species have the same correlator, and so the index $j$ is dropped for $f^{\alpha,\beta}$. Then
\begin{align}
\langle B_{z}(0)B_{z}(t) \rangle = \sum_j  D^2_j(r_j)& \left[ 9 f^{x,x} \left( (u^j_xu^j_z)^2 + (u^j_yu^j_z)^2 \right)\right. \notag\\
&\ \left. + f^{z,z} \left(3u^j_zu^j_z-1\right)^2 \right]. \label{eq:corrEnd}
\end{align}
Assuming that the energy of the nuclear spin state $\vert m_z\rangle$ is $\hbar \omega_{m_z} m_z$, the transverse $f^{x,x}$, $f^{y,y}$ and longitudinal $f^{z,z}$ spin-spin correlation functions have their natural expression in frequency-space with the definition in Eq.\ \eqref{eq:conv}. The relevant spin projections $I_{\alpha}$ for each nucleus are found using their respective operators:
\begin{align}
I_{\alpha} &= \langle n_z \vert \hat I_{\alpha} \vert m_z\rangle
\end{align}
Then in the spectral representation
\begin{gather}
f^{\alpha,\alpha}(I,T,\omega) = \mathscr{F}(f^{\alpha,\alpha}(t)) = \int_{- \infty}^{+ \infty} \langle I_{\alpha}(t) I_{\alpha}(0) \rangle e^{-i \omega t} \diff t \notag \\
\quad = \frac{2 \pi}{Z} \sum_{n,m} e^{-\frac{E_n}{k_B T}} | \langle n_z | \hat I_{\alpha} | m_z \rangle |^2 \delta \left( \tfrac{E_m-E_n}{\hbar} - \omega \right),
\label{eq:transvCorr2}
\end{gather}
where $Z$ is the spin partition function and $E_{m,n}$ are the energies of nuclear spins $m,n$.\cite{Schwabl05} In the high temperature limit where $E_n \ll k_B T$, the eigenstates are equally populated, and
\begin{equation}
f^{\alpha,\alpha}(I,\omega) = \frac{2 \pi}{\text{Tr(\textbf{1})}} \sum_{n,m} \left| \langle n_{z}|\hat I_\alpha|m_z\rangle \right|^2 \delta \left( \tfrac{E_m-E_n}{\hbar} - \omega \right).
\end{equation}
We now make use of the definitions for the $z$ and $x$ spin projections:
\begin{align}
I_z &= \langle n_z \vert \hat I_z \vert m_z\rangle = m_z \langle n_z \vert m_z \rangle \nonumber \\
I_x &= \langle n_z \vert \frac{\hat I^{+} + \hat I^{-}}{2} \vert m_z\rangle,
\end{align}
where
\begin{align}
\hat I^{\pm}|I,m_z\rangle &=\sqrt{I(I+1) - m_z(m_z\pm 1)}|I,m_z\pm1\rangle.
\end{align}
Then the longitudinal correlator is
\begin{equation}
f^{z,z}(I,\omega) = \frac{2 \pi}{\text{Tr(\textbf{1})}} \hat \sum_z \left| \langle m_z|I_z|m_z\rangle \right|^2 \delta(\omega). \label{eq:diagCorr}
\end{equation}
The correlator \eqref{eq:diagCorr} can be computed by noting that a Curie-Weiss prefactor appears due to the relation $\sum_z m_z^2/\text{Tr(\textbf{1})} = I(I+1)/3$. Because the longitudinal correlator is centered at zero energy, it will not contribute to the final integral \eqref{eq:Stau13} as long as $g(\omega=0,\tau,N)=0$ (i.e., the dynamical decoupling pulse sequence is not sensitive to DC fields).
The transverse correlator is
\begin{equation}
f^{x,x}(I,\omega) = \frac{2 \pi}{\text{Tr(\textbf{1})}} \sum_{n,m} \left| \langle n_{z}|\hat I_x|m_z\rangle \right|^2 \delta \left( \tfrac{E_m-E_n}{\hbar} - \omega \right), \label{eq:transvCorr}
\end{equation}
which is non-zero only when $m_z, n_z$ are adjacent energy levels. For the case of spin-1/2 nuclei ($I=1/2$), where the nuclear spins precess at Larmor frequency $\omega_L = \gamma_n B_0$, we evaluate \eqref{eq:transvCorr} as:
\begin{equation}
f^{x,x}(I=1/2,\omega) = \frac{2 \pi}{8} \left( \delta(\omega - \omega_L) + \delta(\omega + \omega_L)\right). \label{eq:transv4}
\end{equation}
The two contributions in \eqref{eq:transv4} represent the Stokes and anti-Stokes lines, equal in the limit $T \rightarrow \infty$.\cite{Schwabl05}

The expression for magnetic field correlation is now
\begin{equation}
\langle B_{z}(0) B_{z}(t) \rangle =  9 f^{x,x} \sum_j  D^2_j(r_j) \left[ (u^j_xu^j_z)^2 + (u^j_yu^j_z)^2 \right], \label{eq:corrEnd2}
\end{equation}
with $f^{x,x}$ given by Eq. \ref{eq:transv4}. By writing $1-(u^j_z)^2 = (u^j_x)^2+(u^j_y)^2$, the geometry-dependent terms can be collected into one factor:
\begin{equation}
\Gamma = \sum_j  D^2_j(r_j) (u^j_z)^2 \left( 1 - (u^j_z)^2 \right), \label{eq:geo}
\end{equation}
which we evaluate in the following section.

\subsubsection{Calculation of the Geometrical Factor}
\label{secB:geoFac}
\begin{figure}[]
  \includegraphics[width=0.75\columnwidth]{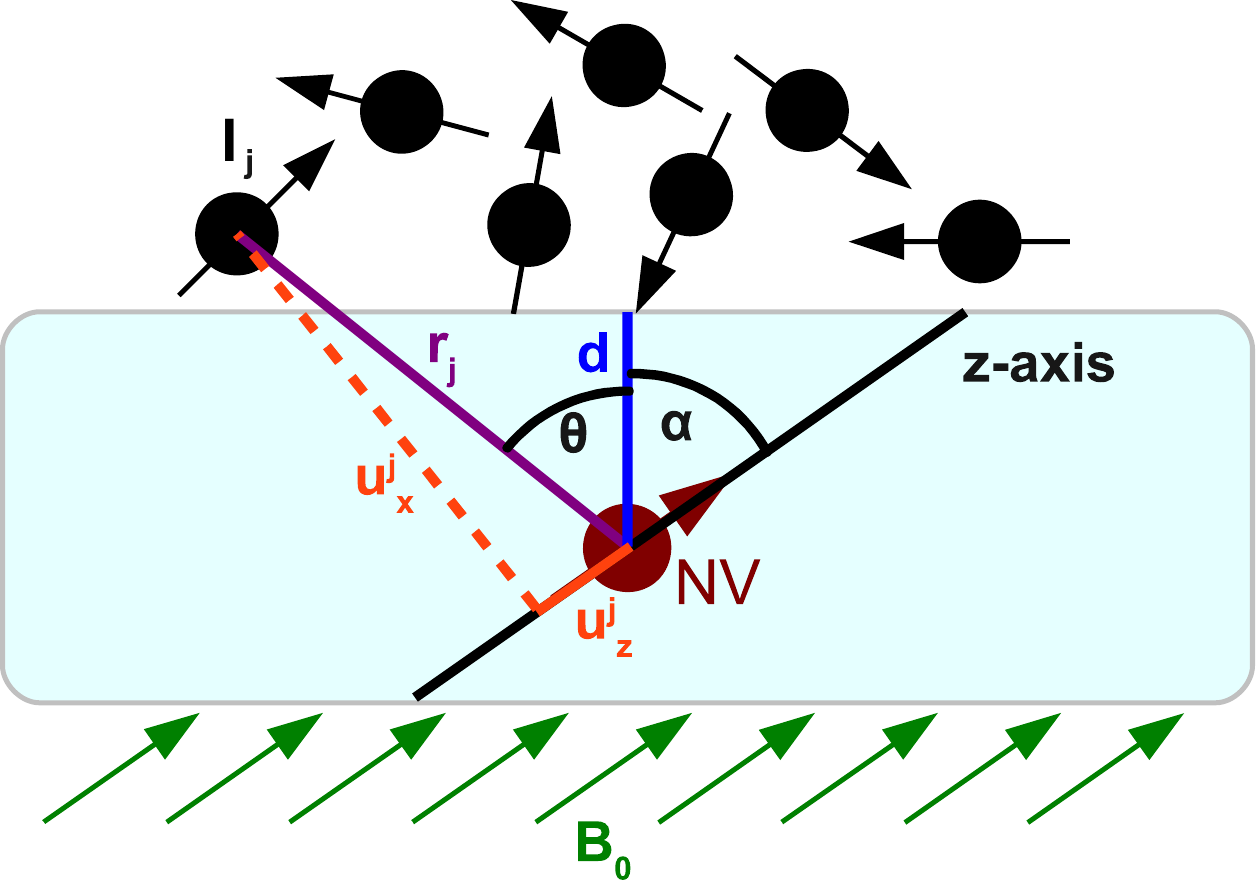}
  \caption{An NV center at depth $d$ below the diamond surface on which resides a sample containing an ensemble of nuclear spins, each with spin vector $I_j$ and position $u^j_x,u^j_y,u^j_z$. The NV axis, and the axis for magnetic quantization, is at angle $\alpha$ with respect to the vector normal to the diamond surface. For purposes of integration across the sample, the spherical coordinates $r,\theta,\phi$ are used. The external magnetic field $B_0$ is assumed to be aligned with the N-V symmetry axis.}
\label{fig:geometry}
\end{figure}
For liquid samples such as immersion oil in which nuclear locations vary on a time scale short compared with the dynamical decoupling sequence length, one can assume a sample of nuclear density $\rho$ continuously distributed on the diamond surface. Then the summation of the geometrical factor \eqref{eq:geo} can be converted to the integral:
\begin{align}
\Gamma &= \rho \int \diff V  \left[ \left( \frac{\mu_0 \hbar \gamma_n}{4 \pi} \right)^2  \frac{(u^j_z)^2(1-(u^j_z)^2)}{r^6} \right] \nonumber \\
&= \rho \left( \frac{\mu_0 \hbar \gamma_n}{4 \pi} \right)^2 \tilde{\Gamma}. \label{eq:geo2}
\end{align}
We evaluated the integral $\tilde{\Gamma}$ using spherical coordinates with the conventions of Fig.\ \ref{fig:geometry}. The polar angle origin $\theta=0$ is defined to be orthogonal to the surface of the diamond, while $\phi$ is the azimuthal angle with arbitrary origin. The NV axis $\mathbf{z}$ points along a direction $\mathbf{z} = \left[ \sin(\alpha)\cos(\beta), \sin(\alpha)\sin(\beta), \cos(\alpha) \right]$. The projection $u_z$ needed for Eq.\ \eqref{eq:geo2} will in general depend on all four angles just introduced. In particular, $u_z = \mathbf{z} \cdot \mathbf{u}_r$, where $\mathbf{u}_r = \left[ \sin(\theta)\cos(\phi), \sin(\theta)\sin(\phi), \cos(\theta) \right]$.    \\
The integral for $\tilde{\Gamma}$ is then
\begin{align}
\tilde{\Gamma} = \int_0^{2\pi} \int_0^{\pi/2} \int_{d_{NV}/\cos(\theta)}^{\infty} \frac{(u_z)^2(1-(u_z)^2)}{r^4} \sin(\theta) \diff r \diff \theta \diff \phi, \label{eq_spherInt}
\end{align}
where $d_{NV}$ is the NV depth below the diamond surface. The sample height is assumed to be semi-infinite, thereby allowing integration of the radial component from the diamond surface to infinity. Other sample geometries can be accomodated with the proper integral limits and choice of coordinate system (i.e., spherical, cylindrical, etc.).
Evaluating the integral produces a simple expression for $\Gamma(d_{NV})$:
\begin{equation}
\Gamma(d_{NV}) = \rho \left( \frac{\mu_0 \hbar \gamma_n}{4 \pi} \right)^2  \left(\frac{\pi \left[8 - 3 \sin^4(\alpha) \right]}{288 d_{NV}^3}   \right). \label{eq:geo2Final}
\end{equation}
The expression is maximal when $\alpha=0$, where $\tilde{\Gamma}(d_{NV}) = \pi/(36 d_{NV}^3)$ However, in most diamond samples, the normal to the surface is aligned along the $[100]$ crystal direction, so that $\alpha = 54.7^{\circ}$. At this angle, $\tilde{\Gamma}(d_{NV}) = 5\pi/(216 d_{NV}^3)$.
With the correlation functions and geometric factors now evaluated, the spectral density can be written as:
\begin{align}
S_B(\omega) &= \langle \left| B_{z}(I=1/2, \omega) \right|^2 \rangle \notag \\ &= \Gamma(d_{NV})
\frac{9 \pi}{4} \left( \delta(\omega - \omega_L) + \delta(\omega + \omega_L)\right). \label{eq:fullB2}
\end{align}
The spectral density can be related to the magnetic field variance from the NMR signal by:
\begin{equation}
S_B(\omega) = \pi B_{\rm{RMS}}^2 \left( \delta(\omega - \omega_L) + \delta(\omega + \omega_L)\right), \label{eq:SBB2}
\end{equation}
where
\begin{align}
B_{\rm{RMS}}^2 & = \frac{9}{4}\Gamma(d_{NV}) \nonumber \\
 & = \rho \left( \frac{\mu_0 \hbar \gamma_n}{4 \pi} \right)^2  \left(\frac{\pi \left[8 - 3 \sin^4(\alpha) \right]}{128 d_{NV}^3}   \right). \label{eq:BRMS}
\end{align}
For NV centers oriented at $\alpha = 54.7^{\circ}$ this simplifies to:
\begin{equation}
B_{\rm{RMS}}^2 = \rho \left( \frac{\mu_0 \hbar \gamma_n}{4 \pi} \right)^2  \left(\frac{5\pi}{96 d_{NV}^3}   \right). \label{eq:BRMS54}
\end{equation}
If the nuclear spin sample on the diamond surface is semi-infinite laterally but not vertically, such as a thin layer between coordinates $z_1$ and $z_2$ above the diamond surface, then Eq.\ \ref{eq:BRMS} can be rewritten as:
\begin{align}
B_{\rm{RMS}}^2 = \rho & \left( \frac{\mu_0 \hbar \gamma_n}{4 \pi} \right)^2  \left(\frac{\pi \left[8 - 3 \sin^4(\alpha) \right]}{128}   \right)\nonumber \\
& \left(\frac{1}{(d_{NV}+z_1)^3}-\frac{1}{(d_{NV}+z_2)^3}\right). \label{eq:BRMS_height}
\end{align}

\subsubsection{The Filter Function $\vert g(\omega,\tau) \vert^2$}
To complete evaluation of the accumulated NV spin phase variance integral \eqref{eq:Stau13} and thus the signal contrast Eq.\ \eqref{eq:contr1}, the filter function $\vert g(\omega,\tau) \vert^2$ must be determined for the dynamical decoupling sequence. For a CPMG or XY8 sequence with $N$ $\pi$-pulses, such as that in Fig.\ 1c, we compute the Fourier transform:
\begin{align}
g(\omega,\tau,N) = \frac{2}{\pi} \sum_{k=- \infty}^{+\infty} & \frac{N \tau (-1)^k}{2k+1} e^{-i\frac{N\tau}{2}\left(\omega - \frac{(2k+1)\pi}{\tau}\right)} \nonumber \\
 & \text{sinc}\left[\frac{N \tau}{2}\left(\omega -\frac{(2k+1)\pi}{\tau}\right) \right]  . \label{eq:filterOmC}
\end{align}
For most purposes, only the first-order terms in Eq.\ \eqref{eq:filterOmC} need to be retained. Additional terms contribute only to higher harmonics, which are not measured in this work. The expansion must include $k=0,-1$ to be symmetric around $\pm \omega$. However, the integral over positive and negative frequencies will be equivalent to twice the integral over positive frequencies as long as $k_{\text{B}}T \gg \hbar \omega_L$.
If the nuclear spin dephasing time is assumed to be infinite, such that the nuclear spin signal can be described by delta functions, we can now obtain a final formula for the signal contrast in the $I=1/2$ case, keeping terms $k=0,-1$:
\begin{widetext}
\begin{align}
C(\tau) \approx \exp \left\{- \frac{2}{\pi^2} \gamma_e^2  B_{\rm{RMS}}^2 (N \tau)^2 \left( \text{sinc}^2 \left[\frac{N\tau}{2} \left(\omega_L - \frac{\pi}{\tau}\right)  \right]\right.\right. & +\text{sinc}^2 \left[\frac{N\tau}{2} \left(\omega_L + \frac{\pi}{\tau}\right)  \right] \nonumber \\ & + \left.\left. 2\ \text{sinc} \left[\frac{N \tau}{2} \left(\omega_L - \frac{\pi}{\tau}\right)  \right] \text{sinc} \left[\frac{N\tau}{2} \left(\omega_L + \frac{\pi}{\tau}\right)  \right] \right) \right\}.
\end{align}
\end{widetext}
The off-resonant terms contribute very weakly to the lineshape and can be ignored, resulting in an approximate formula:
\begin{equation}
C(\tau) \approx \exp \left[-\frac{2}{\pi^2} \gamma_e^2  B_{\rm{RMS}}^2 (N \tau)^2 \text{sinc}^2 \left(\frac{N \tau}{2} \left(\omega_L - \frac{\pi}{\tau} \right)  \right) \right]. \label{eq:finContr}
\end{equation}

\subsection{Nuclear spin dephasing time}
In the previous section, we assumed that the nuclear spin signal could be represented by a delta function, meaning that it has a dephasing time ($T_{2n}^*$) much longer than the length of the NV dynamical decoupling sequence. However, the effective nuclear spin linewidth is broadened due to both dephasing from spin-spin interactions and diffusion through the nanoscale NV interaction volume. In order to take these effects into account, we substitute the delta functions of Eq.\ \eqref{eq:transv4} with normalized Lorentzian functions such that:

\begin{align}
f^{x,x}(I=1/2,\omega) =
\frac{2 \pi}{8}& \left(\frac{1}{\pi} \frac{T_{2n}^{*-1}}{(\omega - \omega_L)^2  + (T_{2n}^{*-1})^2} \right. \nonumber \\
& + \left. \frac{1}{\pi} \frac{T_{2n}^{*-1}}{(\omega + \omega_L)^2 + (T_{2n}^{*-1})^2} \right). \label{eq:transv42}
\end{align}

\noindent As before, we need to compute:
\begin{align}
C(\tau) &= \exp \left(- \frac{\langle \Delta \phi^2(\tau) \rangle}{2} \right) \notag\\ &=\exp \left(- \frac{1}{\pi} \gamma_e^2  B_{\rm{RMS}}^2 \int_{\omega} f^{x,x}(I,\omega) \left| g(\omega,\tau,N)  \right|^2 \diff \omega \right) \label{eq:logCon1}.
\end{align}
Once again, symmetry allows us to simplify the expression using only the positive-frequency component if we multiply the expression by two, leading to:
\begin{widetext}
\begin{equation}
C(\tau) = \exp \left( -\frac{2}{\pi^2} \gamma_e^2  B_{\rm{RMS}}^2 \int_{\omega} \frac{1}{\pi} \frac{T_{2n}^{*-1}}{(\omega - \omega_L)^2 + (T_{2n}^{*-1})^2}  (N \tau)^2 \text{sinc}^2 \left[\frac{N \tau}{2} \left(\omega - \frac{\pi}{\tau} \right)  \right] \diff \omega \right).
\end{equation}
\end{widetext}
It is evident that the integral is a convolution between a Lorentzian $l(\omega)$ and a function $\psi(\omega) \sim \text{sinc}^2(u)$. Using the convolution theorem, the integral can be solved by multiplying the respective Fourier transforms and then taking the inverse Fourier transform of the result. The Lorentzian component is
\begin{equation}
l(\omega) = \frac{1}{\pi} \frac{T_{2n}^{*-1}}{(\omega - \omega_L)^2 + (T_{2n}^{*-1})^2}.
\end{equation}
Its Fourier transform is
\begin{equation}
L(t) = \left(e^{-t T_{2n}^{*-1}-i t \omega_L} H(t)+e^{t T_{2n}^{*-1}-i t \omega_L} H(-t) \right),
\end{equation}
where $H(t)$ is the Heaviside step function. The sinc$^2(u)$ component is
\begin{equation}
\psi(\omega) = (N \tau)^2 \text{sinc}^2 \left[\frac{N \tau}{2} \left(\omega \right)  \right].
\end{equation}
Notice that the frequency offset $\pi/\tau$ has been removed to simplify the Fourier transform. The Fourier transform is
\begin{align}
\Psi(t) = \pi\left[\right.&(t-N\tau) ~\text{sgn}(t-N\tau)\notag\\
&\left.-2 t ~\text{sgn}(t) +(t+N\tau)~\text{sgn}(t+N\tau) \right].
\end{align}
Taking the inverse Fourier transform $K(\omega) = \mathcal{F}^{-1}(L(t)\Psi(t))$, and using the identity $\omega = \pi/\tau$ for the filter function resonance condition, gives the expression:
\begin{widetext}
\begin{multline}
K(\tau) \approx \frac{2 T_{2n}^{*2}}{\left[1 +
   T_{2n}^{*2} \left(\omega_L - \frac{\pi}{\tau}\right)^2\right]^2} \left\{e^{-\frac{N \tau}{
     T_{2n}^{*}}} \left[\left[1 - T_{2n}^{*2} \left(\omega_L - \frac{\pi}{\tau}\right)^2\right] \cos \left[
        N \tau \left(\omega_L - \frac{\pi}{\tau}\right)\right]\right.\right.\\ -
      \left.\left. 2 T_{2n}^{*} \left(\omega_L - \frac{\pi}{\tau}\right) \sin \left[
        N \tau \left(\omega_L - \frac{\pi}{\tau}\right)\right]\right]+
     \frac{N \tau}{T_{2n}^{*}} \left[1 + T_{2n}^{*2} \left(\omega_L - \frac{\pi}{\tau}\right)^2\right] +
      T_{2n}^{*2} \left(\omega_L - \frac{\pi}{\tau}\right)^2 - 1 \right\}.
\label{eq:integr}
\end{multline}
\end{widetext}

\noindent The final expression for signal contrast, including nuclear spin dephasing and again ignoring off-resonant terms in the filter function, is
\begin{equation}
C(\tau) \approx \exp \left( -\frac{2}{\pi^2} \gamma_e^2 B_{\rm{RMS}}^2 K(\tau) \right).
\label{eq:logConFinal2}
\end{equation}

\noindent In practice, experimental determination of whether the nuclear spin $T_{2n}^*$ is long or short relative to the length of the NV dynamical decoupling sequence can be carried out by checking the scaling of the observed contrast dip amplitude and width as a function of $N$ and $\tau$. 

\subsection{Pseudospin Derivation}

An alternative derivation of the signal contrast $C(\tau)$ can be obtained using the pseudospin formalism \cite{Ajoy2015}. The contrast is a product of the pseudo-spin signal $\mathcal S^j$ from each nuclear spin $j$ in the sample on the diamond surface:
\begin{equation}
C(\tau)=\prod_j \mathcal S^j.
\end{equation}
For a CPMG sequence (or XY8) with N pulses, the pseudo-spin signal for nuclear spin $j$ is
\begin{equation}
\mathcal S^j =\!1\!-\!2 \vec\omega^j_0\!\times\!\vec\omega^j_1\sin^2\!\left(\!\frac{\Omega^j_0\tau}{4}\!\right)\sin^2\!\left(\!\frac{\Omega^j_1\tau}{4}\!\right)\frac{\sin^2(\frac{N\alpha^j}{2})}{\cos^2(\frac{\alpha^j}{2})},  \label{eq:SCP}
\end{equation}
where
\begin{align}
 \cos(\alpha^j)=&\cos\left(\frac{\Omega^j_0\tau}{2}\right)\cos\left(\frac{\Omega^j_1\tau}{2}\right) \notag\\
 &-\vec\omega^j_0\cdot\vec\omega^j_1\sin\left(\frac{\Omega^j_0\tau}{2}\right)\sin\left(\frac{\Omega^j_1\tau}{2}\right)
\end{align}
is the effective NV spin rotation angle during one cycle. Here the vectors $\vec \Omega^j_i=\Omega^j_i\vec{\omega^j_i}$ represent the sample nuclear spin Hamiltonians in the two subspaces of the NV electronic spin, i.e., $i$ takes the value of the NV spin state -1, 0, or 1. In the case of nuclear spin-1/2, we have $\vec\omega^j_0=\omega^j_L\hat z$, where $\omega_L$ is the nuclear spin Larmor frequency. On the other hand, $\vec \omega_1^j=\omega^j_L\hat z +\vec A_z^j$, where $\vec A_z^j$ is the dipolar coupling component along the NV z axis. Then the dip in the signal, $\mathcal D^j=1-\mathcal S^j$, can be related to contrast by:
\begin{align}
C& (\tau) = \prod_j \mathcal S^j=\prod_j [1-\mathcal D^j] \nonumber \\
& = \prod_j \left[1- 2(\vec\omega^j_0\!\times\!\vec\omega^j_1)\sin^2\!\left(\!\frac{\Omega^j_0\tau}{4}\!\right)\sin^2\!\left(\!\frac{\Omega^j_1\tau}{4}\!\right)\frac{\sin^2(\frac{N\alpha^j}{2})}{\cos^2(\frac{\alpha^j}{2})}\right]. \label{eq:prod2sum}
\end{align}
The expression can be further simplified in the limit $\omega_L\gg |A_z^j|$, where $\vec A_z^j=A_z^j[\cos\varphi\sin\vartheta,\sin\varphi\sin\vartheta,\cos\vartheta]$. Then, to second order in $A^j_z$, the signal is determined by:
\begin{equation}
\mathcal S^j\approx 1-\frac{2 (A^j_z)^2 \sin ^2(\vartheta)}{\omega _L^2}\frac{ \sin ^4\left(\frac{\omega _L \tau}{4}\right) \sin ^2\left(\frac{N \omega _L \tau}{2}\right)}{ \cos ^2\left(\frac{\omega _L \tau}{2}\right)}.
\end{equation}
For simplicity in the following steps, we define $\kappa_j=A_z^j \sin(\vartheta^j) = (A^j_{zx})^2+(A^j_{zy})^2$. We can also simplify Eq.\ \eqref{eq:filterOmC} using all $k$ values to get
\begin{equation}
\vert g(\omega_L,\tau)\vert^2 =\frac{16}{\omega_L^2} \frac{ \sin ^4\left(\frac{\omega_L \tau}{4}\right) \sin ^2\left(\frac{N \omega_L \tau}{2} \right)}{ \cos ^2\left(\frac{\omega_L \tau}{2}\right)}.
\end{equation}
Then the NV signal contrast from an ensemble of nuclear spins precessing at Larmor frequency $\omega_L$ is
\begin{equation}
C(\tau)=\textstyle\prod_j \left(1-\frac{1}{8}\vert g(\omega_L,\tau) \vert^2 \kappa_j^2\right).
\label{eq:signalsum}
\end{equation}
This product can be reconciled with the exponential form of the previous section in the following manner. First a variance of the effective field is defined as:
\begin{equation}
\langle \kappa^2 \rangle = \frac{1}{n} \sum_{j=1}^n \kappa^2_j.\label{eq:variance}
\end{equation}
The variance is just an average of the individual $\kappa_j^2$ values. If the number of nuclear spins $n$ is large, one can assume that each spin acts like an average spin, and $\kappa_j^2$ can be replaced with $\langle \kappa^2 \rangle$. Then the product simplifies to:
\begin{align}
C(\tau) &= \textstyle\prod_j \left(1-  \frac{1}{8}\vert g(\omega_L,\tau) \vert^2 \kappa_j^2\right)\notag\\& \Rightarrow \left(1-\frac{1}{8}\vert g(\omega_L,\tau) \vert^2 \langle \kappa^2 \rangle\right)^n.
\end{align}
Substitution with Eq.\ \eqref{eq:variance} yields:
\begin{equation}
C(\tau) = \left(1-\frac{1}{8}\vert g(\omega_L,\tau) \vert^2 \frac{1}{n}\sum_j \kappa_j^2\right)^n.
\end{equation}
Note that for large $n$ this is the definition of the exponential. Then
\begin{align}
C(\tau) &= \lim_{n\to\infty} \left(1-\frac{1}{8}\vert g(\omega_L,\tau) \vert^2\frac{1}{n}\sum_j \kappa_j^2\right)^n \notag\\ &= \exp \left(-\frac{1}{8}\vert g(\omega_L,\tau) \vert^2 \sum_j \kappa_j^2\right).\label{eq:exp1}
\end{align}
The term $\sum_j \kappa_j^2$ can converted into an integral of the form $\int \rho(\vec r) \kappa^2(\vec r) \diff^3r$ and integrated over the sample. Since $A_z$ represents the frequency shift from dipolar coupling, one can show from the definition of $\kappa$ that:
\begin{equation}
\sum_j \kappa_j^2 = 9 \gamma^2_e \sum_j D^2_j(r_j) (u^j_z)^2 \left(1 - (u^j_z)^2 \right) = 4 \gamma^2_e B_{\rm{RMS}}^2. \label{eq:kappa}
\end{equation}
This along with the approximated expression of the filter function finally allows Eq.\ \eqref{eq:exp1} to be written as:
\begin{align}
C(\tau) & \approx \exp \left(-\frac{1}{2}\gamma^2_e\vert g(\omega_L,\tau) \vert^2 B_{\rm{RMS}}^2\right) \nonumber \\
 & = \exp \left(-\frac{2}{\pi^2} \gamma^2_e (N \tau)^2 \text{sinc}^2 \left(\frac{N \tau}{2} \left(\omega_L - \frac{\pi}{\tau} \right)\right) B_{\rm{RMS}}^2\right). \label{eq:exp2}
\end{align}
Importantly, the expression \eqref{eq:exp2} for contrast exactly matches that given in Eq.\ \eqref{eq:finContr}, showing the equivalence of the two calculational approaches presented here.

\renewcommand\thefigure{\thesection.\arabic{figure}}
\setcounter{figure}{0}
\section{Estimated Proton Nanoscale NMR Linewidth Calculated from Correlation Time}

The NV NMR protocol detects a nuclear spin signal via the dipole-dipole interaction, which makes it extremely sensitive to changes in nuclear spin position. As a consequence of the strong distance dependence of dipolar coupling, nuclei diffusing in a liquid on the diamond surface move in and out of the nanoscale sensing volume very quickly, which limits the interaction time between the NV and nuclear spin. As a result, the nanoscale NMR linewidth is broadened. This is in contrast to conventional NMR detection via an inductive coil surrounding the sample, in which the nuclei can be fully contained within the sensing volume and changes in nuclear position have little effect on the signal. 

We assume that the interaction between the NV and nuclear spin lasts for a characteristic correlation time, $\tau_d$, and that the probability of finding the particles interacting drops off exponentially in time. By taking the Fourier transform, this behavior produces a Lorentzian lineshape $L(\omega)$ typically written as:
\begin{equation}
L(\omega) = \frac{1}{\pi} \frac{\tau_d}{1+\omega^2 \tau_d^2}.
\end{equation}

\begin{figure}[]
\includegraphics[width=0.65\columnwidth]{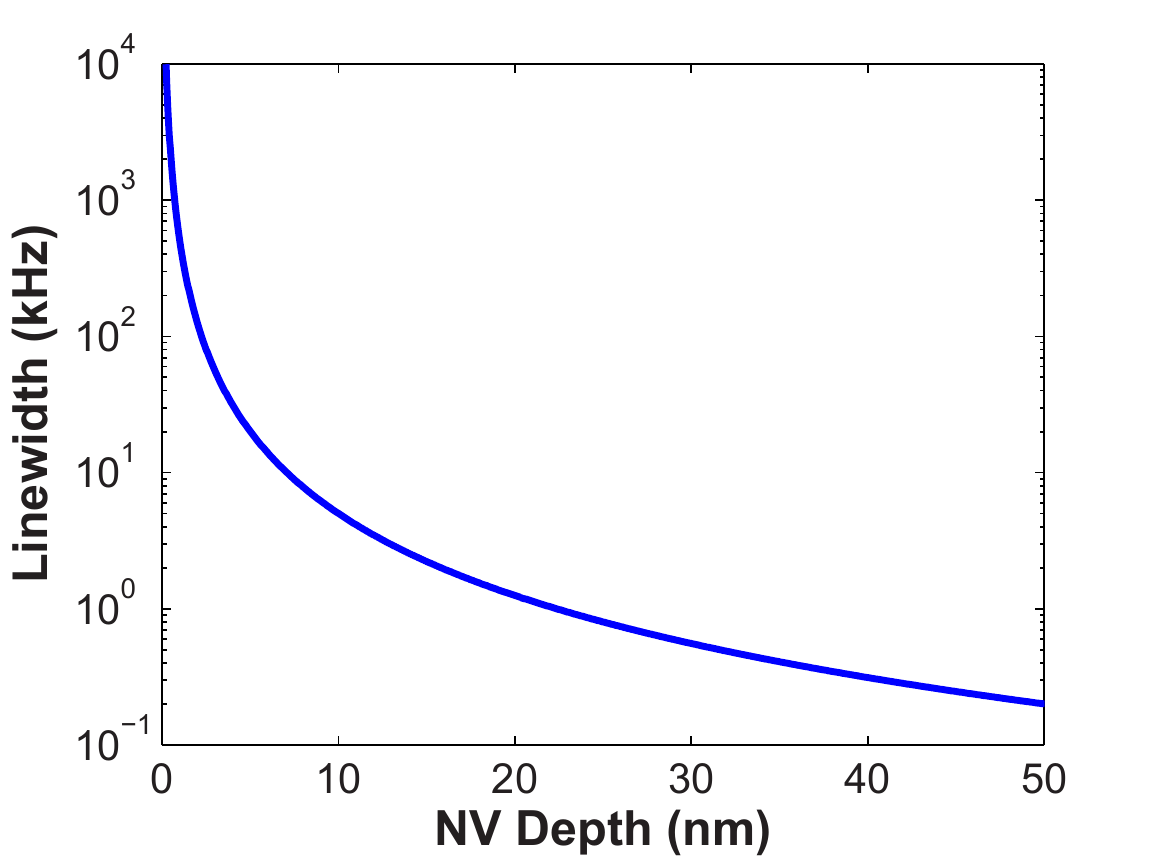}
\caption[Estimate of proton nanoscale NMR linewidth for NV NMR experiment]{Estimate of proton nanoscale NMR linewidth as a function of NV depth, for immersion oil on the diamond surface.}
\label{fig:estimated_linewidth}
\end{figure}

\noindent This can also be written in a standard Lorentzian form:
\begin{equation}
L(\omega) = \frac{1}{\pi} \frac{1/\tau_d}{\omega^2+1/\tau_d^2}.
\end{equation}
The full width at half maximum (FWHM) is then $2/\tau_d$.

The translational diffusion correlation time for two spins in three dimensions (in our case the immobile NV and diffusing nuclei in molecules in the sample) can be related to molecular geometries and diffusion coefficients by \cite{hubbard1963nuclear, Potenza1}:
\begin{equation}
\tau_d = \frac{d^2}{D_{av}},
\end{equation}
where $d$ is the distance of closest approach between the two spins and $D_{av}$ is the average of the diffusion coefficients for the two spins. Since the NV center is immobile, we can assume that its diffusion coefficient is zero. The distance of closest approach is the NV depth, $d_{NV}$. Then the correlation time becomes:
\begin{equation}
\tau_d = \frac{2 d_{NV}^2}{D_{nuc}},
\label{eq:corrtime}
\end{equation}
where $D_{nuc}$ is the diffusion coefficient of the molecules in the sample carrying the nuclear spins.

Low-fluorescent immersion oil is typically composed of liquid polybutadiene mixed with smaller amounts of paraffins and carboxylic acid esters \cite{oil_patent}. In one example of an immersion oil with kinematic viscosity $\nu = 450$ cSt \cite{oil_patent}, the polybutadiene component has an average molecular weight of 1600 g/mol. The hydrodynamic radius of the molecule is on the order of $r \sim 1$ nm \cite{Fetters1}, and the density is $\rho \sim 0.9$ g/mL. The dynamic viscosity is then
\begin{equation}
\eta = \rho \nu = 0.405 \rm{~cP}.
\end{equation}
We use this viscosity as an approximation for the similar immersion oil employed in our experiment. Using the Stokes-Einstein relationship
\begin{equation}
D = \frac{k_B T}{6 \pi \eta r}
\end{equation}
gives a diffusion coefficient $D_{oil} \approx 5 \times 10^{-13}$ m$^2$/s.

Figure \ref{fig:estimated_linewidth} plots the estimated nanoscale NMR linewidth for immersion oil as a function of NV center depth calculated using equation \ref{eq:corrtime}. The estimated NMR linewidth is $\sim5$ kHz for a $\sim10$ nm deep NV center, while the broadest NMR linewidth we expect to see in the measurements performed in this work is $\sim30$ kHz for a $\sim4$ nm deep NV center. Consequently, we expect that the NV NMR detection bandwidth is much broader than the sample's NMR linewidth (i.e., the infinite $T_{2n}^*$ approximation is valid) for nearly every measurement, excepting measurements with long pulse sequence durations on the shallowest NV centers.

\bibliography{depth_references}

\end{document}